\documentclass[apjl,iop]{emulateapj}
\usepackage{color} 
\usepackage{multirow}
\usepackage{rotating}
\usepackage{framed} 
\usepackage{booktabs}
\usepackage{hyperref} 
\usepackage[all]{hypcap} 
\usepackage{amsmath}

\newcommand{\lsim}{\mathrel{\hbox{\rlap{\lower.55ex\hbox{$\sim$}} \kern-.3em \raise.4ex \hbox{$<$}}}}
\newcommand{\gsim}{\mathrel{\hbox{\rlap{\lower.55ex\hbox{$\sim$}} \kern-.3em \raise.4ex \hbox{$>$}}}}
 \tighten

 \newcommand{\BE}{\begin{equation} }
 \newcommand{\EE}{\end{equation} }
 \newcommand{\BEA}{\begin{eqnarray} }
 \newcommand{\EEA}{\end{eqnarray} }
 \newcommand{\msun}{M_\sun}
 \newcommand{\pc}{\,{\rm pc}}
  
 \newcommand{\Mpc}{\,{\rm Mpc}} 
  \newcommand{\Gpc}{\,{\rm Gpc}}

 \newcommand{\Teff}{T_{\rm eff}}
 
 \newcommand{\yr}{\,{\rm yr}}
  \newcommand{\Gyr}{\,{\rm Gyr}}

 \newcommand{\BEn}{\begin{enumerate}}
 \newcommand{\EEn}{\end{enumerate}}
 \newcommand{\BI}{\begin{itemize}}
 \newcommand{\EI}{\end{itemize}}
 \newcommand{\En}{\mathcal{E}}
 \newcommand{\rate}{\mathcal{R}}
 \newcommand{\fsky}{f_{\rm sky}}
 \newcommand{\rategal}{\Gamma} 
 \newcommand{\igammafunc}{Q}
 \newcommand{\kms}{\,{\rm km \, s}^{-1}}
 \newcommand{\Mbh}{M_{\rm BH}}
 \newcommand{\tobs}{t_{\rm obs}}

 \def\r{\mathbf{r}}
 \def\v{\mathbf{v}}

\begin{document}
\title{Multiple Tidal Disruptions as an Indicator of Binary Super-Massive Black Hole Systems}
\author{Christopher\ Wegg\altaffilmark{1} and J.~Nate\ Bode}
\affil{Theoretical Astrophysics, California Institute of
Technology, MC 350-17, 1200 East California Boulevard,
Pasadena, CA 91125}
\altaffiltext{1}{ \href{mailto:wegg@tapir.caltech.edu}{wegg@tapir.caltech.edu}}  
\shorttitle{MULTIPLE TIDAL DISRUPTIONS INDICATE MBH BINARIES} 
\shortauthors{Wegg~\&~Bode}

\begin{abstract} 
We find that the majority of systems hosting multiple tidal disruptions are likely to contain hard binary SMBH systems, and also show that the rates of these repeated events are high enough to be detected by LSST over its lifetime. Therefore, these multiple tidal disruption events provide a novel method to identify super-massive black hole (SMBH) binary systems with parsec to sub-parsec separations. The rates of tidal disruptions are investigated using simulations of non-interacting stars initially orbiting a primary SMBH and the potential of the model stellar cusp. The stars are then evolved forward in time and perturbed by a secondary SMBH inspiraling from the edge of the cusp to its stalling radius. We find with conservative magnitude estimates that the next generation transient survey LSST should detect multiple tidal disruptions in approximately $3$ galaxies over $5$ years of observation, though less conservative estimates could increase this rate by an order of magnitude.
\end{abstract}
\keywords{ black hole physics --- galaxies: kinematics and dynamics --- galaxies: evolution ---
 galaxies: nuclei}
\section{Introduction} 
\label{introduction}

Stars with radius $r_\star$, and mass $M_\star$, which pass within the tidal disruption radius $r_t \sim r_\star (\Mbh/M_\star )^{1/3}$ of a super-massive black hole (SMBH) of mass $\Mbh$,
 will be ripped apart by tidal forces. In the case of sun-like stars,
 \BE
 r_t\approx1.2 r_s M_8^{-2/3}\,, \label{eqn:rt}
 \EE 
where $r_s$ is the Schwarzschild radius, and $M_8$ is $\Mbh/10^8\msun$. Therefore, when $\Mbh \gsim 10^8 \msun$ the Schwarzschild radius lies outside $r_t$ and any sun-sized star would be swallowed whole. Below this critical black hole mass the star's debris is launched on orbits which span an energy range $\Delta E  \approx  G \Mbh r_\star / r_t^2$ \citep{Rees:1988}. This energy range is large compared to the energy of the highly elliptic initial orbit, and hence half the material will be unbound while half will fall back onto the black hole. For main sequence stars the fall back rate declines as $t^{-5/3}$ \citep{Phinney:1989}. This fall back rate is initially super-Eddington for the canonical 10\% accretion efficiency \citep{Evans:1989p871}, but it is unclear whether the disk adjusts to lower its accretion rate or a radiatively driven outflow results \citep{Strubbe:2009,Lodato:2010}.

Galaxies harboring an isolated SMBH at their center are expected to quickly clear a `loss cone' of orbits whose angular momenta about the black hole are low enough that their peribothra lie inside $r_t$. At this point tidal disruptions are predicted at a rate $\sim10^{-4}-10^{-5} \yr^{-1}$ as stars diffuse into the loss cone \citep{Magorrian:1999p478,Wang:2004,Donley:2002p682}.
The majority of candidate tidal disruptions thus far have been found though X-ray \citep[e.g.][]{Donley:2002p682} or UV surveys \citep[e.g.][]{Gezari:2008p205}. This is expected, as can be seen by modeling the tidal disruption as a thick disk emitting as a black body with luminosity $L_{\rm edd}$, temperature $\Teff$, and initially extending to $r_t$. In reality the disk will expand outwards on a viscous timescale and the initial super Eddington rate could launch an outflow. Ignoring these complications, however, gives \citep{Ulmer:1999} 
\BE
\Teff\sim3.7\times10^5 M_8^{1/12}\left({M_\star \over M_\sun}\right)^{-1/6}  \left({r_\star \over r_\sun}\right)^{-1/2}\,{\rm K}\,,
\EE 
and the spectrum peaks in the extreme UV. Despite being in the Rayleigh-Jeans tail of this flux, optical transient surveys such as the Palomar Transient Factory (PTF), the Panoramic Survey Telescope and Rapid Response System (Pan-STARRS) and the Large Synoptic Survey Telescope (LSST) provide the prospect of finding many more tidal disruptions because of their unprecedented combination of high cadence and depth. It is expected that LSST will detect a striking $\sim 100-3000 \yr^{-1}$ \citep{Strubbe:2009}, where the major uncertainty is how the luminosity from the uncertain super-Eddington phase of the tidal disruption is included.

In this letter we calculate the rates of multiple tidal disruptions from the same merging SMBH binary system, and show that the detection of multiple tidal disruptions from a single galaxy likely indicates the galaxy hosts a SMBH binary with a parsec to subparsec separation. 
Our results are summarized in table \ref{tab:mergerates}.

\section{Simulations}
\label{simulations}

Two mechanisms for enhanced rates of tidal disruptions in close SMBH binary systems have been considered in the literature, the Kozai effect \citep{Ivanov:2005p402} and chaotic 3-body orbits \citep{Chen:2009}. We have extended this work by performing a series of restricted 3-body scattering experiments including two additional key aspects: the evolution of the binary, and the non-Keplarian stellar potential. Our simulations are described below, where we use $G=1$ throughout. 

In considering the stellar potential, it is important to have a stellar distribution consistent with a central SMBH. Thus, the primary SMBH of mass $M_1$ was placed in a Tremaine-model (also known as the Dehnen-model) cusp \citep{Tremaine:1994,Dehnen:1993} whose density is given by 
\BE
\rho(r)=\frac{\eta}{4\pi}\frac{1}{r^{3-\eta} (1+r)^{1+\eta}} \, .
\label{eq:etarho}
\EE
The advantage of this model is that it self-consistently describes a finite mass of stars distributed around a central black hole together with their stellar potential
\begin{align}
\Psi(r) &= \frac{1}{\eta-1}\left[1 -  \frac{r^{\eta-1}}{(1+r)^{\eta-1}}\right] + \frac{\mu}{r} \, , & \eta \neq 1\, , \\
&=\ln(1+1/r) +\frac{\mu}{r} \, , & \eta=1\,,\notag
\end{align}
where $\mu$ is the ratio of $M_1$ to the total stellar mass in the cusp. Throughout this work we use $\mu=0.5$. Since the stellar mass is normalized to unity this results in the primary's mass being half the cusp's stellar mass. A Bachcall-Wolf cluster corresponds to $\eta=1.25$.

In this way our simulations for a given $\eta$ and $q$ then depend only on the tidal disruption radius $r_t$. The scaling of these parameters to real galaxies is described later by equations \ref{eq:rbullet} and \ref{eq:rtrc}. 

Each star's initial position was chosen according to equation \ref{eq:etarho} with a random orientation. To pick each star's velocity, we numerically tabulated the distribution function $f(\En)$ and then picked an isotropic velocity from the required distribution, $4\pi f(\En) v^2 \, dv = 4\pi f(\En) \sqrt{2(\Psi(r)-\En)}\,d\En$, using the rejection method.

\begin{figure}[htp]
\centering
\includegraphics[width=\linewidth]{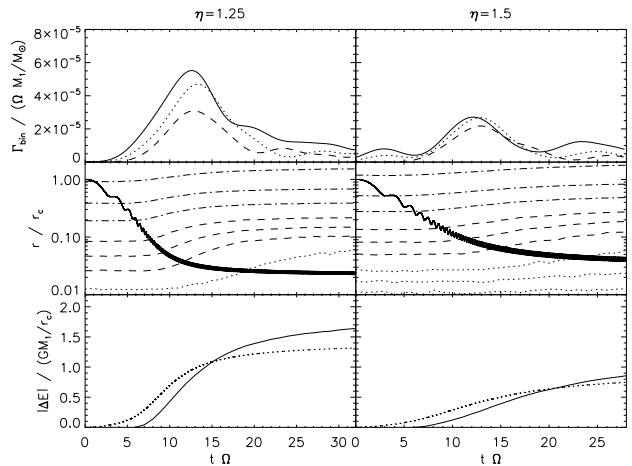}
\caption{Results for mass ratio $q=0.1$. The left hand panels show an $\eta=1.25$, Bahcall-Wolf cusp, the right hand panels an $\eta=1.5$ cusp. The upper panels shows tidal disruption rate, $\Gamma_{\rm bin}$, for tidal disruption to cusp radius ratios $r_t/r_c=(9,7,5)\times10^{-7} $ in solid, dotted and dashed lines respectively. The distribution of disruption times have been kernel smoothed with a Gaussian of width $\sigma=2$. The middle panels shows the evolution of the binary separation as a solid line and radii enclosing $0.1$, $0.2$, $0.4\%$ of the stellar mass in dotted, $1$, $2$, $4\%$ in dashed and $10$, $20$, $40\%$ in dash-dot lines. The lower panel shows the evolution of the energy of the binary as the solid line and the stars as the dotted. In a fully self-consistent evolution these would lie on top of each other. The simulations are scaled by $\Omega\equiv(2GM_1/r_c^3)^{1/2}$.} 
\label{fig:q0p10resultsmatrix}
\end{figure}

The stars are initially on an orbit consistent with the primary SMBH and the stellar potential, however their orbits are perturbed by the secondary SMBH whose orbit we evolve with time approximately following an inspiral dominated by dynamical friction. To model this inspiral, the secondary SMBH is initially given an eccentricity of $0.1$ and a binary separation equal to the cusp radius, $r_c$. It is then migrated inwards on a path governed by
\BE
\frac{d \v}{dt} = -  \frac{\left[\mu(1+q)+M_\star(<r) \right]}{r^3}\r - \omega_{\rm df} \v \,
\EE
where $M_\star(<r)$  is the stellar mass interior to $r$, $q\le1$ is the binary mass ratio and
\BE 
\omega_{\rm df}= \frac{4\pi\log\!\Lambda\, q \: \mu \: \rho(< v)}{v^3}
\EE
characterizes the dynamical friction \citep{Binney:2008}. Here $\rho(<\!v)$ is the density of stars at $r$ with velocity less than $v$. We have used a Coulomb logarithm that begins at $\log \Lambda\approx4$, but which smoothly decreases to zero at the stalling radius calculated by \cite{Sesana:2008}. The functional form of the decrease was chosen to approximate the rate of shrinkage caused by the ejection of stars during our scattering experiments. For instance, in the lower panels of figure \ref{fig:q0p10resultsmatrix} we plot the change in stellar and binary energies. If the functional form of $\log \Lambda$ had been chosen perfectly the two would lie on top of each other.

\begin{figure}[htp]
\centering
\includegraphics[width=\linewidth]{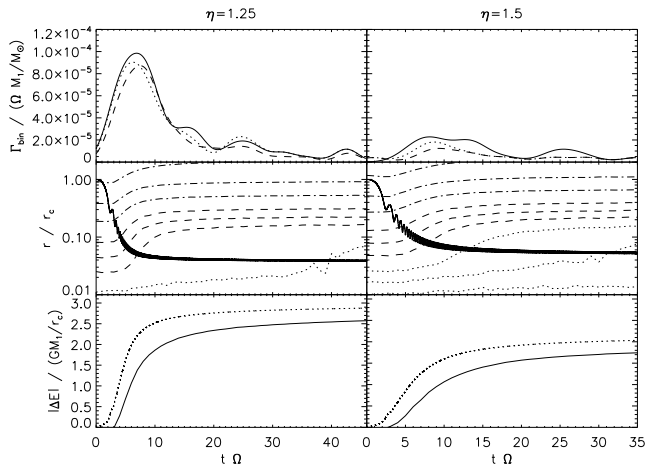}
\caption{Results as described in figure \ref{fig:q0p10resultsmatrix} for $q=0.3$.}
\label{fig:q0p30resultsmatrix}
\end{figure}
 
\begin{figure}[htp]
\centering
\includegraphics[width=\linewidth]{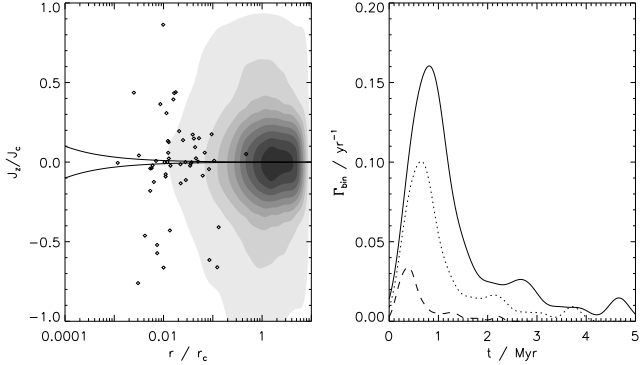}
\caption{Plots of our $\eta=1.25$, $q=0.3$ simulation. Left panel shows the stars that are tidally disrupted for $r_t/r_c=5\times10^{-7}$ as a function of their in initial radius and $z$-component of angular momentum normalized to the circular angular momentum at that radius, $J_z/J_c$.  The Kozai wedge is plotted together with the overall stellar density. A large fraction of the disrupted stars lie well outside the Kozai wedge indicating that these are chaotic orbits \citep[cf.][]{Chen:2009}. The contours show the initial stellar distribution, each is evenly spaced in density. The right hand panel shows the rates from the same simulation scaled using the relations in equations \ref{eq:rbullet} and \ref{eq:rtrc} for $M_1=10^8\msun$ in solid, $5\times10^7\msun$ in dotted and $10^7\msun$ dashed lines.} 
\label{fig:f3new}
\end{figure}

To perform our scattering experiments we have implemented the adaptive symplectic integrator described in \citet{Preto:1999}, with a timestep which varies as $\Delta t \sim U^{-1}$ where $U$ is the potential energy. With this choice of timestep the integrator has the desirable property that it reproduces exactly Keplarian orbits in Keplarian potentials, allowing Kozai resonances to be correctly reproduced since the spurious precession frequently found in other algorithms is absent. In addition, this integrator is well suited for this problem since it has been shown to correctly reproduce the highly eccentric orbits required for a star to be tidally disrupted \citep{Peter:2009}. In our simulations stars are considered disrupted when they pass within the tidal disruption radius of either hole appropriate for a sun-like star. We find the vast majority of stars are disrupted by the primary SMBH. 

It should be noted that for the duration of the scattering experiments the stellar potential is assumed to be centered on the primary SMBH and is not allowed to vary with time. These assumptions are made for simplicity, and will be relaxed in future studies. 

The results of our scattering experiments and the resultant disruption rates, $\Gamma_{\rm td}$, are shown in figures \ref{fig:q0p10resultsmatrix} and \ref{fig:q0p30resultsmatrix} for mass ratios $q=0.1$ and $q=0.3$, respectively. The plots are scaled by $\Omega\equiv(2GM_1/r_c^3)^{1/2}$.

We find that the majority of tidal disruptions are due to the type of chaotic orbits described by \citet{Chen:2009} as opposed to the Kozai effect discussed by \citet{Ivanov:2005p402}. This is demonstrated in the left hand panel of figure \ref{fig:f3new}. The reason for this is twofold: precession of the orbits of the stars in the non-Keplarian potential destroys the secular Kozai effect for the majority of orbits, and the few stars closely bound to the primary whose precession rate is lower have Kozai timescales longer than our simulation.

Our rates are lower than those discussed by \citet{Chen:2009} largely because we have considered less steep cusps. This both reduces the number of stars that can be disrupted as the binary hardens, and increases the orbital timescale at the hardening radius. Both effects reduce the rate of disruptions. In addition we have considered the binary evolution which \citet{Chen:2009} did not, although this has a smaller effect 

To apply our simulations to physical galaxies, we use the fits from \cite{Merritt:2009} to the inner regions of ACS Virgo Cluster galaxies \citep{Cote:2004}. For power-law galaxies these give \footnote{D. Merritt personal communication. From fitting to figure 2 of \citet{Merritt:2009}.} 
\BE
r_\bullet = 22\,(M_1 / 10^8 \msun)^{0.55}\pc \, , 
\label{eq:rbullet} 
\EE
where $r_\bullet$ is defined such that the stellar mass interior to $r_\bullet$ is $2 M_1$ and $M_1$ is the mass of the SMBH. Matching this to the Tremaine model such that the central densities are equal gives $r_c=r_\bullet$ for $\mu=0.5$, used in our simulations. The ratio $r_t/r_c$ is
\BE
r_t/r_c = 4.9\times10^{-7} (M_1 / 10^8 \msun)^{0.22}\, .
\label{eq:rtrc}
\EE
With these scalings, our simulations for $\eta=1.25$ and $q=0.3$ are shown in figure \ref{fig:f3new}.

\section{Observable Tidal Disruptions}
\label{observabledisruptions}

The absolute magnitude of an individual disruption is likely to depend on complex details with considerable uncertainties in modeling quantities such as the SMBH mass, the SMBH spin and the geometry of the disruption \citep{Strubbe:2009}. 

We instead derive a simple empirical estimate of the volume accessible by comparison to \citet{Gezari:2008p205}. Two luminous optical events coincident with UV flares  were discovered in $\sim2.9\,\rm{deg}^2$. Their spectra and light curves were consistent with tidal disruption events, making this their most likely explanation. Their redshifts were $z=0.33$ and $z=0.37$, giving extinction corrected (but not K-corrected) absolute $g$-band magnitudes of $-17.7$ and $-18.9$ \citep{Gezari:2010p188}. Requiring that these two cases be $2$ mags brighter than the $25.0$ $g$-band limit of LSST
gives maximum redshifts of detection of $z=0.27$ and $z=0.43$, respectively. The $2$ magnitude buffer better ensures a convincing light curve, which would display the characteristic fast rise and decay of a tidal disruption. Based on these numbers we choose $z=0.35$ as the limit for LSST. 

There is also only a small range of SMBH masses which needs to be considered. Because SMBHs of mass greater than $10^8\msun$ can't tidally disrupt stars (equation \ref{eqn:rt}, ignoring SMBH spin), and SMBHs of mass less than $10^7\msun$ are significantly less luminous, (particularly if the super-Eddington phase is neglected), and so can only be observed in a much smaller volume, we restrict our analysis to SMBHs with masses between $10^7\!-\!10^8\msun$.

\section{Rates of Single Tidal Disruptions}
\label{singledisruptions}
We now calculate the rate of tidal disruptions observable by LSST both for systems with isolated SMBHs and systems with SMBH binaries. 

In the case of isolated SMBHs, the rate of tidal disruptions observed by LSST will be
\BE
\rate_{\rm single}^{\rm (td)} = \fsky  \int \frac{dN}{d\Mbh} V_{\rm c}(\Mbh) \rategal_{\rm td}(\Mbh) \, d\Mbh \label{eqn:Rtdexact}
\EE
where $\fsky$ is the fraction of the sky covered by LSST, $\rategal_{\rm td}$ is the rate of tidal disruptions per galaxy, $V_c$ is the total comoving volume over which a tidal disruption is observable and $dN/d\Mbh$ is the black hole mass function. We have used the black hole mass function \citep{Aller:2002}
\BE
\frac{dN}{d\Mbh} = c \left( \frac{\Mbh}{\Mbh^\star} \right)^{-\alpha} e^{-\Mbh/\Mbh^\star} \, , 
\EE
with the parameters $c=3\times10^{-11} \msun^{-1}\Mpc^{-3}$, $\Mbh^\star=1.1\times10^8\msun$ and $\alpha=0.95$\ \citep[values derived by][scaled to $H_0=71\kms \Mpc^{-1}$]{Aller:2002}. The rate of tidal disruptions per galaxy is highly uncertain, and so we parameterize, $\rategal_{\rm td}=\gamma\times10^{-5}\yr^{-1}$, scaling to the observationally motivated constant rate per galaxy of \citep{Donley:2002p682} independent of $\Mbh$. We also assume $V_{\rm c}$ is independent of $\Mbh$ and is $10.7\Gpc^3$, corresponding to our redshift limit of  $z=0.35$ with the assumption that $H_0=71\kms \Mpc^{-1}$. Using these approximations and $\fsky\approx 0.5$, the rate of tidal disruptions detected by LSST in galaxies containing isolated SMBHs (equation \ref{eqn:Rtdexact}) is predicted to be
\BEA
\rate_{\rm single}^{\rm (td)} &\sim& V_{\rm c} \fsky \rategal_{\rm td} \int_{10^7 \msun}^{10^8 \msun} \frac{dN}{d\Mbh} \, d\Mbh \notag \\
&=& 300\gamma\,\yr^{-1} \, .
\EEA

Now consider systems hosting binary SMBHs. The rate of disruptions will be
\begin{align}
\rate_{\rm bin}^{\rm (td)} &=  \fsky \int  V_{\rm c} \rategal_{\rm bin}(M_1,q,t) R_{\rm merge}(M_1,q)\, dq  \, dM_1\, dt \, ,
\end{align}
where $R_{\rm merge}(M_1,q)\,dq\,dM_1 $ is the rate of mergers per unit comoving volume for binary SMBHs with primary mass between $M_1$ and $M_1+dM_1$ and with mass ratio between $q$ and $q+dq$. The quantity $\int \rategal_{\rm bin}(M_1,q,t) \, dt$ is the total number of tidal disruptions in a merger and was linearly interpolated from the simulations in figures \ref{fig:q0p10resultsmatrix} and \ref{fig:q0p30resultsmatrix} together with equations \ref{eq:rbullet} and \ref{eq:rtrc}. 

Over the narrow range of redshift and primary mass accessible we approximate 
\BE
R_{\rm merge}(M_1,q) = C \frac{dN}{dM_1} F(q) \, ,
\EE
where we have assumed the mass ratio distribution of SMBH binaries follows the local galaxy merger mass ratio distribution given by \citet{Stewart:2009},
$F(q)=q^{0.25} (1-q)^{1.1}$.
The normalization constant $C$ was chosen to be $0.05\,\Gyr^{-1}$, so that we we reproduce the simulated local merger rate\footnote{M. Volonteri personal communication. From data in figure 2 of \citet{Volonteri:2009}.} of SMBHs with primary SMBH mass between $10^7\msun$ and $10^8\msun$ and $q>0.05$ of approximately $9\times 10^{-5}\Mpc^{-3}\Gyr^{-1}$. 

 Using these approximations, then, if all galaxies have an $\eta=1.25$ cusp we can expect LSST to detect 
\begin{align}
\rate_{\rm bin}^{\rm (td)} \sim& V_{\rm c} \fsky C \notag \\  
& \times \int\limits_{10^7\msun}^{10^8\msun}\,dM_1 \int\limits_{0.05}^{0.5} \,dq \int \,dt \frac{dN}{dM_1} F(q)\rategal_{\rm bin}(M_1,q,t)  \notag   \\
=& 10\, \yr^{-1} \, ,
\label{eq:ratebintd}
\end{align}
where we have limited the mass ratio to $q<0.5$ since our simulations assume that the secondary has been stripped of stars when it reaches the cusp.

\section{Rates of Multiple Tidal Disruptions}
\label{multipledisruptions}
\begin{table*}[t]
  \centering

  \begin{tabular}{|cccccccccc|}
  \tableline
  \tableline
 \multirow{2}{*}{Survey} & \multirow{2}{*}{mag limit} &&\multirow{2}{*}{$z_{\rm lim}$} && \multirow{2}{*}{ $f_{\rm sky}$} & \multirow{2}{*}{ $\rate_{\rm single}^{\rm (td)}/\yr^{-1}$}&\multirow{2}{*}{$\rate_{\rm bin}^{\rm (td)}/\yr^{-1}$} &\multirow{2}{*}{$N_{\rm single}^{\rm (multi)}$} & \multirow{2}{*}{$N_{\rm bin}^{\rm (multi)}$}  \\
   & &&  &&  &  &  &  &\\
        \tableline
  PTF 	& $21.0$ ($g$-band)		&& $0.06$ 	&& $0.2$  		& $0.7 \,\gamma $ 	& $0.02$ 	& $0.9\times10^{-4}\,\gamma^2$		 	& $0.007$		\\ 
  PAN-Starrs (MDS) & $25.0$ ($g$-band)	&& $0.35$ 	&& $10^{-3}$ 	& $0.7 \,\gamma $ & $0.02$ 	& $0.004\,\gamma^2$ 			& $0.007$		\\
  LSST 	& $25.0$ ($g$-band) 		&& $0.35$ 	&& $0.5$ 		& $300 \,\gamma $ & $10$ 	& $0.03\,\gamma^2$ 			& $3$		\\
  \tableline
  \end{tabular}
      \caption{Summary of rates of tidal disruptions for three current and upcoming transient surveys. Symbols and calculation are described in the text. The numbers $N_{\rm single}^{\rm (multi)}$ and $N_{\rm bin}^{\rm (multi)}$ are for an observation time of $\tobs=5\yr$ and scale roughly as $\tobs^2$. }

  \label{tab:mergerates}
\end{table*}

We now calculate the rate of multiple tidal disruptions in systems containing isolated SMBHs. Over a period of observing $t_{\rm obs}$ the total number of tidal disruptions will follow a Poisson distribution with mean $t_{\rm obs} \Gamma_{\rm td}$. The probability of observing multiple tidal disruptions from a single galaxy is therefore $1 - \igammafunc(2,t_{\rm obs} \Gamma_{\rm td})$, where $\igammafunc $ is the incomplete gamma function.  

Then the expected number of isolated SMBHs exhibiting multiple tidal disruptions during $t_{\rm obs}$ is
\begin{align}
N_{\rm single}^{\rm (multi)}& = V_c \fsky \int\limits_{10^7\msun}^{10^8\msun}\frac{dN}{d\Mbh} \left[ 1 - \igammafunc(2,t_{\rm obs} \Gamma_{\rm td}) \right] \, d\Mbh  \notag \\
&\sim 0.03 \gamma^2 \left(t_{\rm obs}/5\yr\right)^2  \, .
\end{align}

Similarly, the expected number of multiple tidal disruptions observed from binary SMBHs  is
\BE
N_{\rm bin}^{\rm (multi)} = \fsky \int V_{\rm c} \left[1 - \igammafunc(2,\Gamma_{\rm bin} t_{\rm obs}) \right] R_{\rm merge} \, dM_1\, dt \, dq \, .
\EE
Using the same approximations used in estimating equation \ref{eq:ratebintd} we find over an observation time, $t_{\rm obs}=5\yr$, the expected number of close binary SMBHs exhibiting multiple tidal disruptions observable by LSST to be
\begin{align}
N_{\rm bin}^{\rm (multi)} \sim &V_{\rm c} \fsky C \int_{10^7\msun}^{10^8\msun}\,dM_1 \int_{0.05}^{0.5} \,dq \notag \\
& \times \int \,dt \frac{dN}{d\Mbh} F(q) \left[1 - \igammafunc(2,\Gamma_{\rm bin}(M_1,q,t) t_{\rm obs}) \right]   \notag \\
=&3 \, ,
\end{align}
where the quantity in square brackets was calculated from our simulations together with equations \ref{eq:rbullet} and \ref{eq:rtrc}.

Towards the upper end of the range $10^7-10^8\msun$ $\Gamma_{\rm bin} t_{\rm obs} \sim0.5$ for major mergers. This indicates that the majority of close SMBH binaries with primaries in the upper end of this range could potentially be identified using multiple disruptions. The expression above broadly scales as $ t_{\rm obs}^2$ but because $\Gamma_{\rm bin} t_{\rm obs} \sim0.5$ for some systems this is only approximate.

\section{Discussion}
\label{discussion}

We have estimated the enhanced rate of tidal disruptions from SMBH binaries, and shown that if a system exhibiting multiple tidal disruptions is observed then in our fiducial model it is $\sim100$ times more likely to be a close SMBH binary than an isolated SMBH system. It has also been shown that the upcoming transient survey LSST is likely to detect several systems with multiple disruptions during a $5$ year observation period. 

Once a double tidal disruption is detected these galaxies would be expected to have a steady tidal disruption rate, with further events on a human timescale.  In addition, a double disruption makes the system immediately worthy of detailed study and monitoring to estimate the properties of the predicted binary black hole.

We have also shown that in our fiducial model approximately $3\%$ of all tidal disruptions occur in binaries. Other signatures may identify tidal disruptions that occurred in binaries. These include possible spectroscopic signatures, morphology or kinematics indicating a recent major merger, or interruption of the tidal disruption flare on a binary orbital timescale \citep{Liu:2009p389}.

These results depend strongly on several quantities. In particular, the tidal disruption rates are largely determined by the number of stars in the central regions of the galaxy, which, in turn, depends on the cusp profile and the size of the cusp. In this sense, multiple tidal disruptions are also diagnostic of cusp profiles. We have assumed Bahcall-Wolf type cusps for our rates. Moreover, the rates of multiple tidal disruptions from isolated SMBHs depend on the square of the uncertain tidal disruption rate, for which we have adopted a conservative fiducial rate of $10^{-5}\yr^{-1}$. 

All our numbers scale by the uncertain detection volume, which could be significantly higher than we have assumed. Recently \citet{vanVelzen:2010} found two candidate disruptions with absolute $g$-band magnitudes $-20.3$ and $-18.3$. If the event with magnitude $-20.3$ was representative of the higher black hole mass disruptions where our binary induced disruptions typically occur, then LSST could detect disruptions of this type to $z\sim0.7$ increasing our predicted rates by approximately an order of magnitude.


\section{ Acknowledgments }
We gratefully acknowledge useful discussions with Sterl Phinney, Annika Peter and Andrew Benson. We also thank David Merritt for providing his fits to ACS Virgo data and Marta Volonteri for calculating the SMBH merger rates. 

Support for this work was provided by NASA BEFS grant NNX-07AH06G.

\bibliographystyle{astroads}
\bibliography{tidaldisruptionabbrv}

\end{document}